\documentclass{article}
\usepackage{hhline, pbox, array} 
\usepackage{graphicx, bm} 
\usepackage[caption=false]{subfig}
\usepackage{color}
\usepackage{gensymb}
\usepackage{verbatim}
\usepackage{fullpage}
\usepackage{authblk}
\usepackage{hyperref}

\graphicspath{ {Figures_ARXIV//} }
\begin{document}

\title{\textbf{Suppression of axial growth by boron incorporation in GaAs nanowires grown by self-catalyzed molecular beam epitaxy}}

\author[1]{Suzanne Lancaster}
\author[2]{Heiko Groiss}
\author[1]{Tobias Zederbauer}
\author[1]{Aaron M. Andrews}
\author[1]{Donald MacFarland}
\author[1]{Werner Schrenk}
\author[1]{Gottfried Strasser}
\author[1,3]{Hermann Detz}

\affil[1]{Center for Micro- and Nanostructures, Institute for Solid State Electronics, TU Wien, 1040 Vienna, Austria}
\affil[2]{Center of Surface and Nanoanalytics, Johannes Kepler University Linz, 4040 Linz, Austria}
\affil[3]{Central European Institute of Technology, Brno University of Technology, 612 00 Brno, Czech Republic}
\affil[]{email: suzanne.lancaster@tuwien.ac.at}
\maketitle
\vspace{10pt}

\begin{abstract}
The addition of boron to GaAs nanowires grown by self-catalyzed molecular beam epitaxy was found to have a strong effect on the nanowire morphology, with axial growth greatly reduced as the nominal boron concentration was increased. Transmission electron microscopy measurements show that the Ga catalyst droplet was unintentionally consumed during growth. Concurrent radial growth, a rough surface morphology and tapering of  nanowires grown under boron flux suggest that this droplet consumption is due to reduced Ga adatom diffusion on the nanowire sidewalls in the presence of boron. Modelling of the nanowire growth puts the diffusion length of Ga adatoms under boron flux at around 700-1000nm. Analyses of the nanowire surfaces show regions of high boron concentration, indicating the surfactant nature of boron in GaAs.
\end{abstract}

%
%
%
%
%

\section{Introduction}

Heterostructured nanowires (NWs) offer interesting prospects as the key building blocks of future electronic and optoelectronic devices. In particular, bottom-up III-V NWs such as GaAs offer the advantages of a direct bandgap, and the possibility of growing heterostructures both radially \cite{lauhon2002epitaxial} and axially \cite{paladugu2007novel}, while still relying on established processing technology. The growth of materials with a large lattice mismatch is faciliated in NWs due to the relaxation of strain in two dimensions \cite{caroff2009insb}. At the same time, much recent work has focused on the residual or engineered strain which remains in the nanostructures and can be used as an additional parameter (alongside size and composition) to manipulate material properties such as the emission wavelength \cite{hocevar2013residual, treu2013enhanced} or electron mobility \cite{liu2015coreshell}. Investigating strain in nanowires is particularly interesting for the design of radial heterostructures. 

In order to induce strain in a GaAs nanowire, ternary alloys with two group III or group V elements can be used. While alloys with larger lattice constants than GaAs, such as In$_x$Ga$_{1-x}$As, are well-researched, few traditional III-As candidates exist with a smaller lattice constant than GaAs. The addition of N to GaAs nanowires has been demonstrated in radial GaAs/GaAs$_{1-y}$N$_y$ heterostructures \cite{araki2013growth}, which leads to strain in the NWs and modifies the bandgap \cite{Chen2014origin}, and can lead to further investigations on the quaternary Ga$_{1-x}$In$_x$As$_{1-y}$N$_y$. 

In contrast, it has been hypothesised that small amounts of B lead only to small changes in the GaAs bandgap \cite{gupta2000molecular}, while still allowing for the addition of strain. In that respect, it is interesting to investigate the ternary alloy B$_x$Ga$_{1-x}$As, where the theoretical lattice constant should range from that of BAs, 4.78\AA \cite{wentzcovitch1986theory}, to that of GaAs, 5.65\AA, for strain engineering applications. However, research on this alloy is sparse, owing to the difficulties of epitaxial growth, including surface segregation \cite{dumont2003surface}, rough surface morphologies \cite{gupta2000molecular}, and unintentional doping due to antisite defects \cite{detz2017growth}. 

Nanowires offer an interesting platform for the growth of B$_x$Ga$_{1-x}$As and for elucidating the effects of molecular beam epitaxy (MBE) growth of  III-V materials with B, as well as for characterization of the optical, electrical and structural properties of the material. While many studies exist on the mixing of group V species (e.g. As + P, As + Sb) \cite{dheeraj2008zinc, zhang2013self, ren2016new}, group III elements such as In and Al generally incorporate on the correct lattice site and do not pose a problem for NW growth. However, from planar growth it is clear that this is not the case with B$_x$Ga$_{1-x}$As. As there is currently no literature on the growth of III-V alloys incorporating B in nanostructures, here we report the growth of GaAs/(B)GaAs nanowires and discuss the effects of boron on nanowire morphology, and the implications for the MBE growth of boron-containing III-V alloys, both in planar geometry and in nanostructures.

\begin{table*}[!t]
\resizebox{\textwidth}{!}{%
\begin{tabular}{|c||c|c|c|c|c|}
\hline
Sample &  B cell temp (\degree C) &  Equivalent SL & GaAs stem & Growth under & AlGaAs growth  \\
 & &  B content (at \%) & growth (mins) & B flux (mins) &  time (mins) \\
\hline
A1 & - & 0 & 60 & - & -\\
\hline
A2 & - & 0 & 21 & - & -\\
\hline
B & 1725 & 0.22  & 18 & 42 & - \\
\hline
C & 1750 & 0.26 & 18 & 30 & - \\
\hline
D & 1775 & 0.31 & 18 & 42 & -\\
\hline
E & 1800 & 0.36 & 18 & 30 & -\\
\hline
F & 1800 & 0.36 & 18 & 30 & 12\\
\hline
\end{tabular}}
\caption{\label{table1}Growth times and parameters for the samples discussed in this paper}
\end{table*}

\begin{figure*}[!b]
\subfloat[]{\label{SEM3} \includegraphics[width=0.3\textwidth]{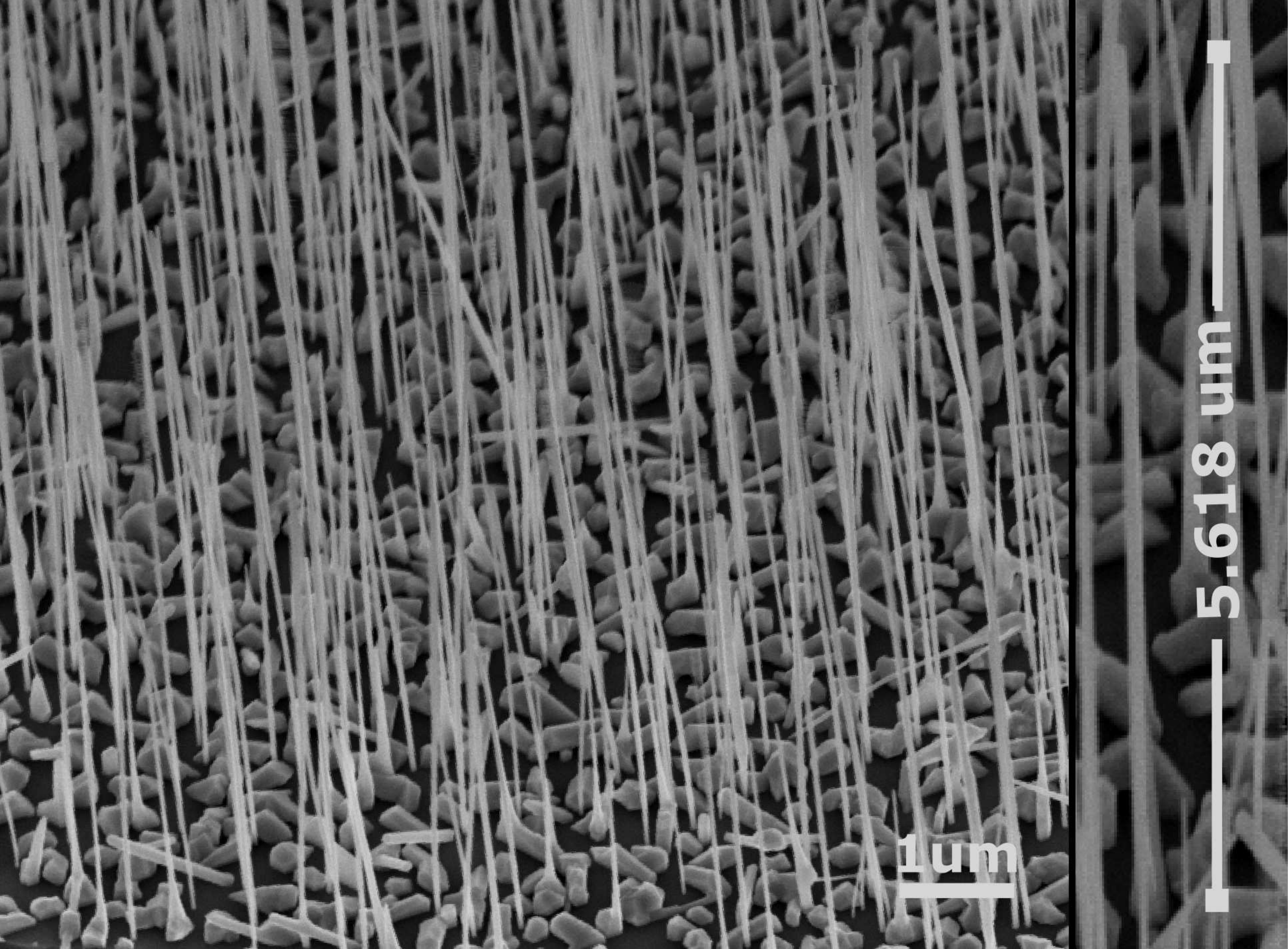}} \hfill
\subfloat[]{\label{SEM1} \includegraphics[width=0.3\textwidth]{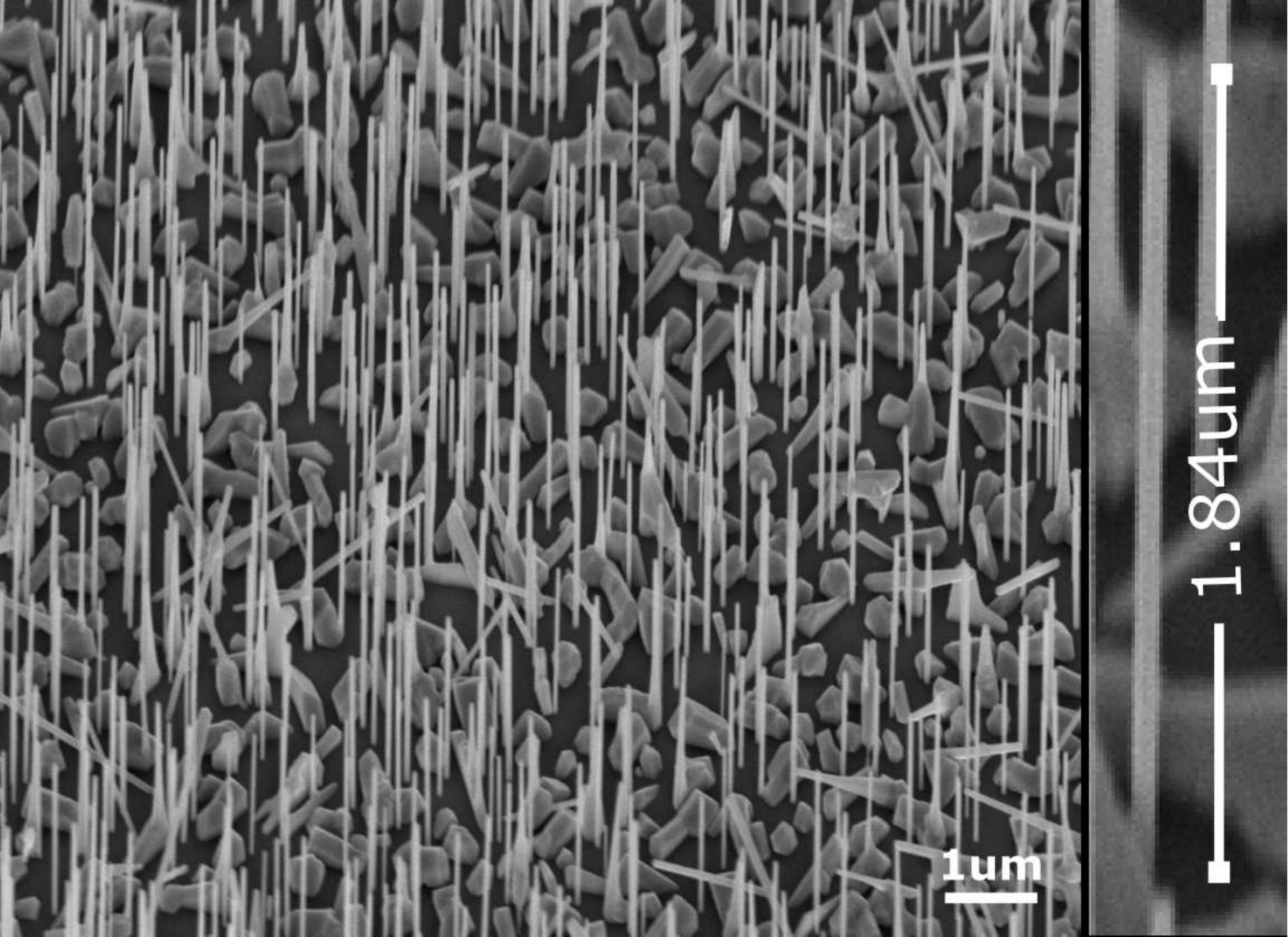}} \hfill
\subfloat[]{\label{SEM2} \includegraphics[width=0.3\textwidth]{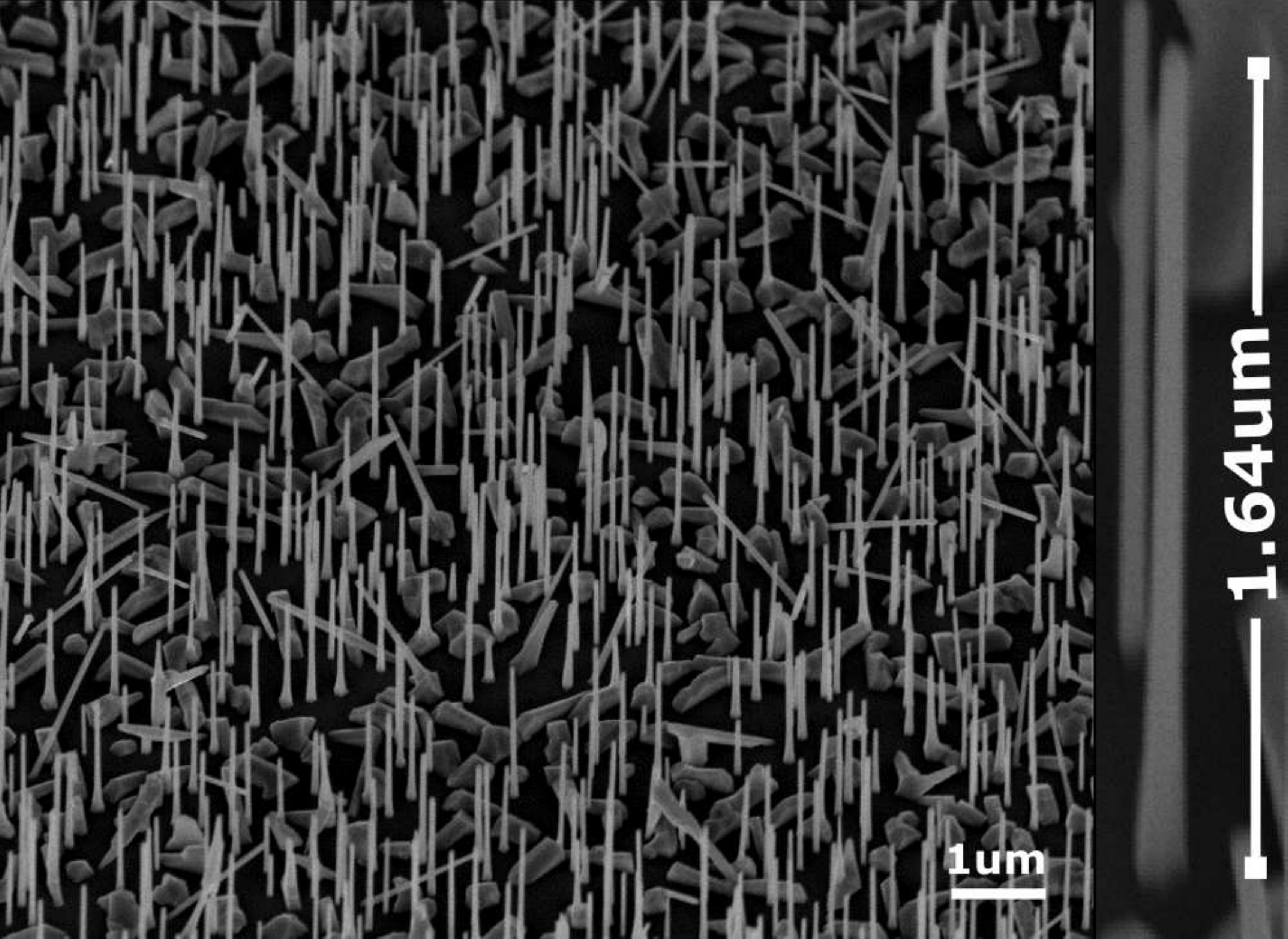}}
\captionsetup{font=footnotesize}
\caption{\label{SEM} Representative SEM images of nanowires from samples grown under  \protect\subref{SEM3} no boron flux (sample A1), \protect\subref{SEM1} the lowest boron flux (sample B) and \protect\subref{SEM2} the highest boron flux (sample E), taken at 45\degree stage tilt}
\end{figure*}

\section{Growth \& Characterization}
All the nanowires described here were grown on Si(111) via a self-catalyzed MBE technique \cite{jabeen2008self}. The whole process was performed at a substrate temperature of 630$\degree$C, measured by a pyrometer calibrated for GaAs. Ga and As were supplied from elemental sources, using an equivalent layer growth rate of 0.1 $\mu$m/h for GaAs and a V/III ratio $\sim$ 7. After the Si (111) substrates were stabilized at the growth temperature, the Ga and As shutters were opened simultaneously for the nucleation of Ga droplets and the subsequent vapor-liquid-solid (VLS) growth of initial GaAs nanowire stems. This initial growth phase was performed without boron to avoid any possible influence on the nucleation behavior. Except for the GaAs reference samples, Boron flux was supplied from a water-cooled high-temperature effusion cell after 18 min of VLS growth. Powdered B of 6N purity was used as the charge material. The flux was determined by the cell temperature setpoint, which was varied between 1725 and 1800$\degree$C, while all other growth parameters were kept the same. Finally, growth was terminated by closing all shutters simultaneously and the sample cooled down under residual As$_4$ flux. The samples discussed are summarised in table \ref{table1}; the control GaAs samples (samples A1 \& A2) were grown for 21 and 60 mins respectively, while samples grown under boron flux were grown for a total of 48 or 60 minutes. The percentage of boron measured for planar GaAs/B$_x$Ga$_{1-x}$As superlattice samples grown at each temperature and with the same planar equivalent growth rate are given as a guide for the boron incorporation. 

A final sample (sample F) was overgrown with an Al$_{0.42}$Ga$_{0.58}$As shell. In this case, after the (B)GaAs growth, the sample was exposed to a pure As$_4$ flux for 10 minutes, after which an Al$_{0.42}$Ga$_{0.58}$As layer was grown for 12 minutes at a substrate temperature of 500 \degree C. Finally, cooldown followed under residual As$_4$ flux as for the previous samples. 

To characterise the grown nanowires, scanning electron microscopy (SEM) measurements were performed in a Zeiss NEON 40xB system. Representative SEM images of as-grown NWs with and without boron are shown in figure \ref{SEM}. We can see that while NWs were obtained under all growth conditions, the morphology of the NWs changes significantly for samples grown under boron flux, with a greatly reduced wire length and more tapering visible in samples grown under boron flux. This tapering suggests a smaller adatom diffusion length on the NW sidewalls. For measurement of individual wire dimensions, NWs were dispersed in isopropanol and drop-casted on Si/SiO$_2$ carrier wafers.

\begin{figure*}[!t]
\begin{center}
	\subfloat[]{\includegraphics[width=0.45\textwidth]{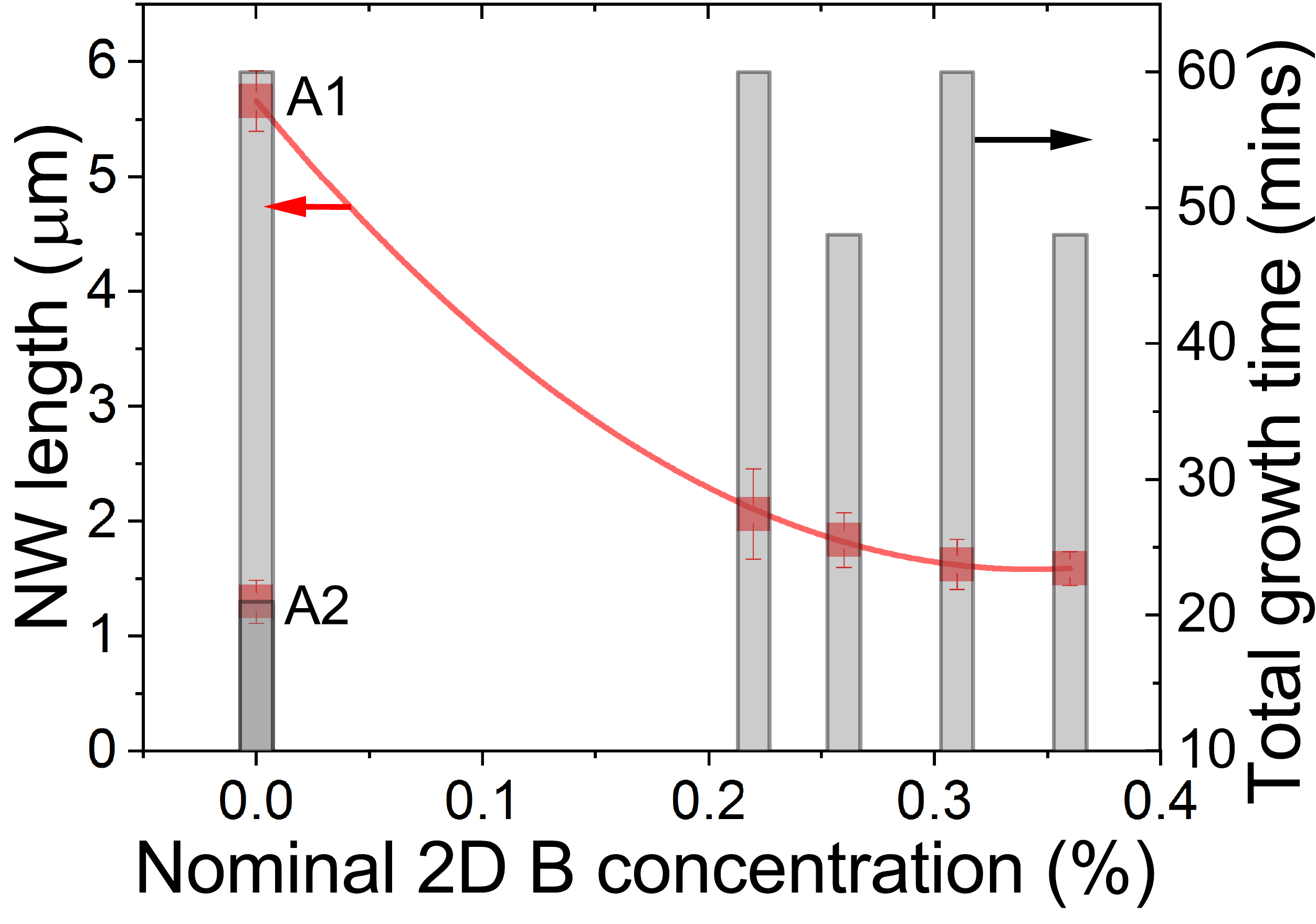}\label{lengthstats}}
	\hfill
	\subfloat[]{\includegraphics[width=0.45\textwidth]{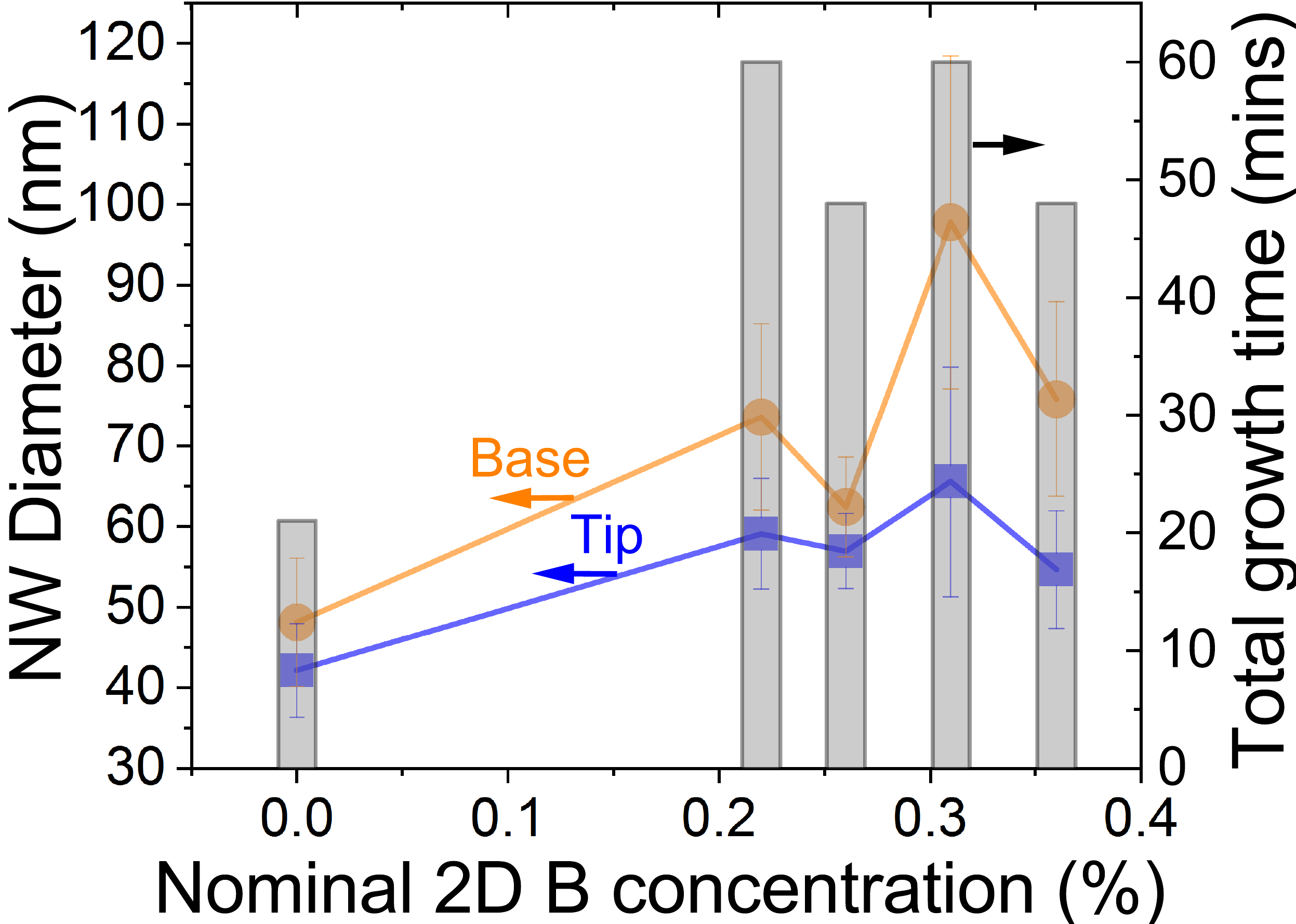}\label{diamstats}}
	\\
	\subfloat[]{\includegraphics[width=0.45\textwidth]{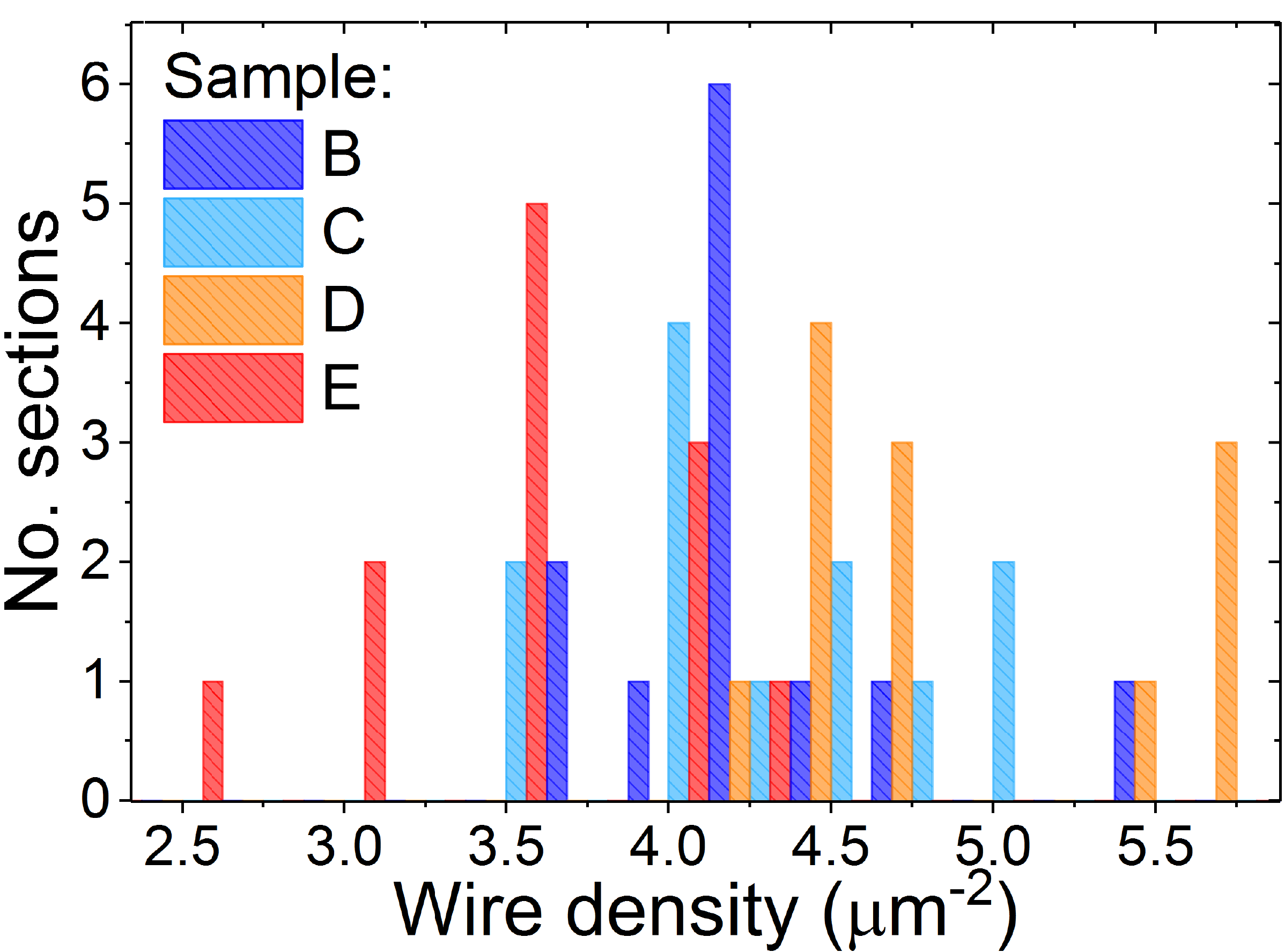}\label{densitystats}}
	\captionsetup{font=footnotesize}
	\caption{\protect\subref{lengthstats} Lengths of control samples and samples grown under B flux (red circles). Total growth time under B flux is indicated on the right axis (grey bars). \protect\subref{diamstats} Average diameter at NW tip (blue line) and base (orange line) against nominal boron concentration. Total growth time is indicated on the right axis (grey bars). \protect\subref{densitystats} Histogram of wire densities of samples grown under B flux, measured on different substrate regions, with a section size of 6$\mu m^2$.}
	\label{stats}
\end{center} 
\end{figure*}

Using the same drop-casting method, some nanowires were transferred onto holey carbon grids with a mesh size of 300$\mu$m, and subsequent TEM analysis was performed on a JEOL JEM-2200FS system using the high resolution (HR) TEM and scanning (STEM) mode. For HRTEM images the wires were tilted to a \textless110\textgreater  zone axis of the GaAs zinc blende (ZB) lattice. Along this direction the ABC stacking of the lattice is clearly visible and allows a simple differentiation between the ZB and the wurtzite (WZ) lattices. High-angle annular dark field (HAADF) and bright field (BF) detectors were used for imaging in the STEM mode. The chemical composition measurements and verification of the B incorporation was done with energy dispersive X-ray spectroscopy (EDXS). 

\begin{figure}[!b]
\captionsetup[subfigure]{labelformat=empty}
\begin{center}
	\subfloat[][]{
		\includegraphics[width=0.6\textwidth]{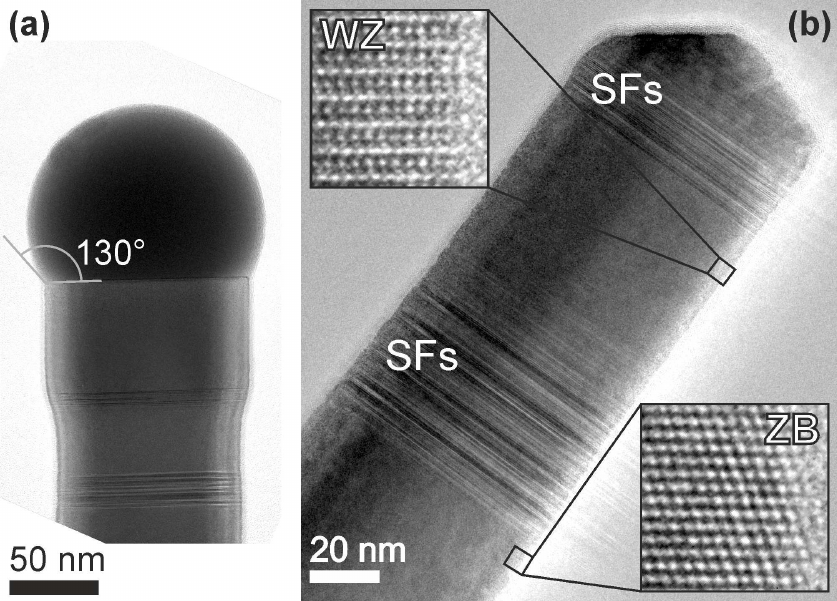}
		\label{droplet_intact}
	}
	\captionsetup{font=footnotesize}
	\caption{HRTEM images of droplets taken after growth. \protect\subref{droplet_intact} Droplet seen intact after growth of pure GaAs. The contact angle of the droplet is indicated. (b) Crystallised droplet on nanowire grown under B flux. Stacking faults and wurtzite crystal structure can be observed under the droplet.}
	\label{droplets}
\end{center} 
\end{figure}

\begin{figure*}[!t]
\captionsetup[subfigure]{labelformat=empty}
\begin{center}
	\subfloat[]{
		\includegraphics[width=0.31\textwidth]{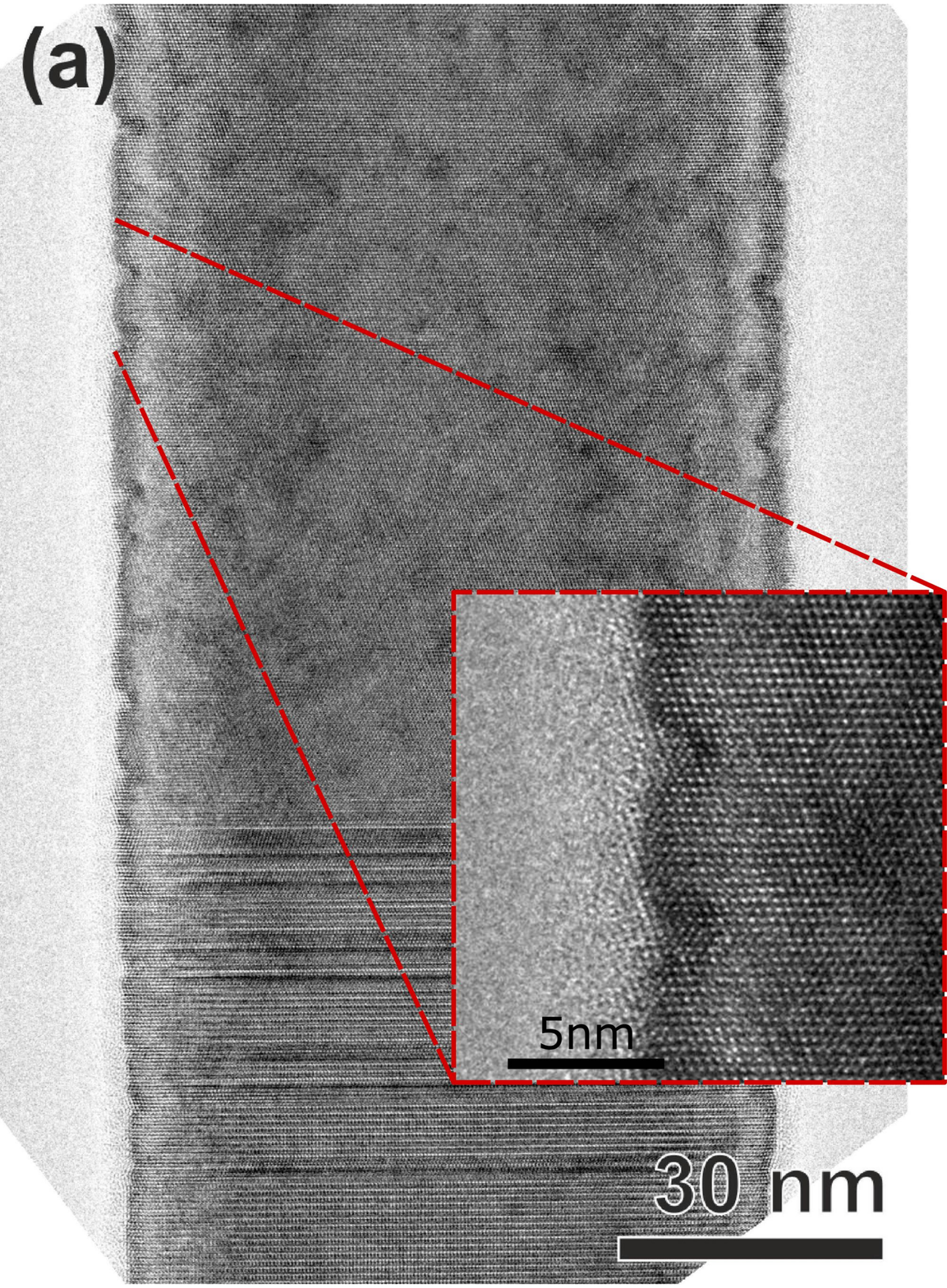}
		\label{HRTEM_380}
	}
	\subfloat[]{
		\includegraphics[width=0.305\textwidth]{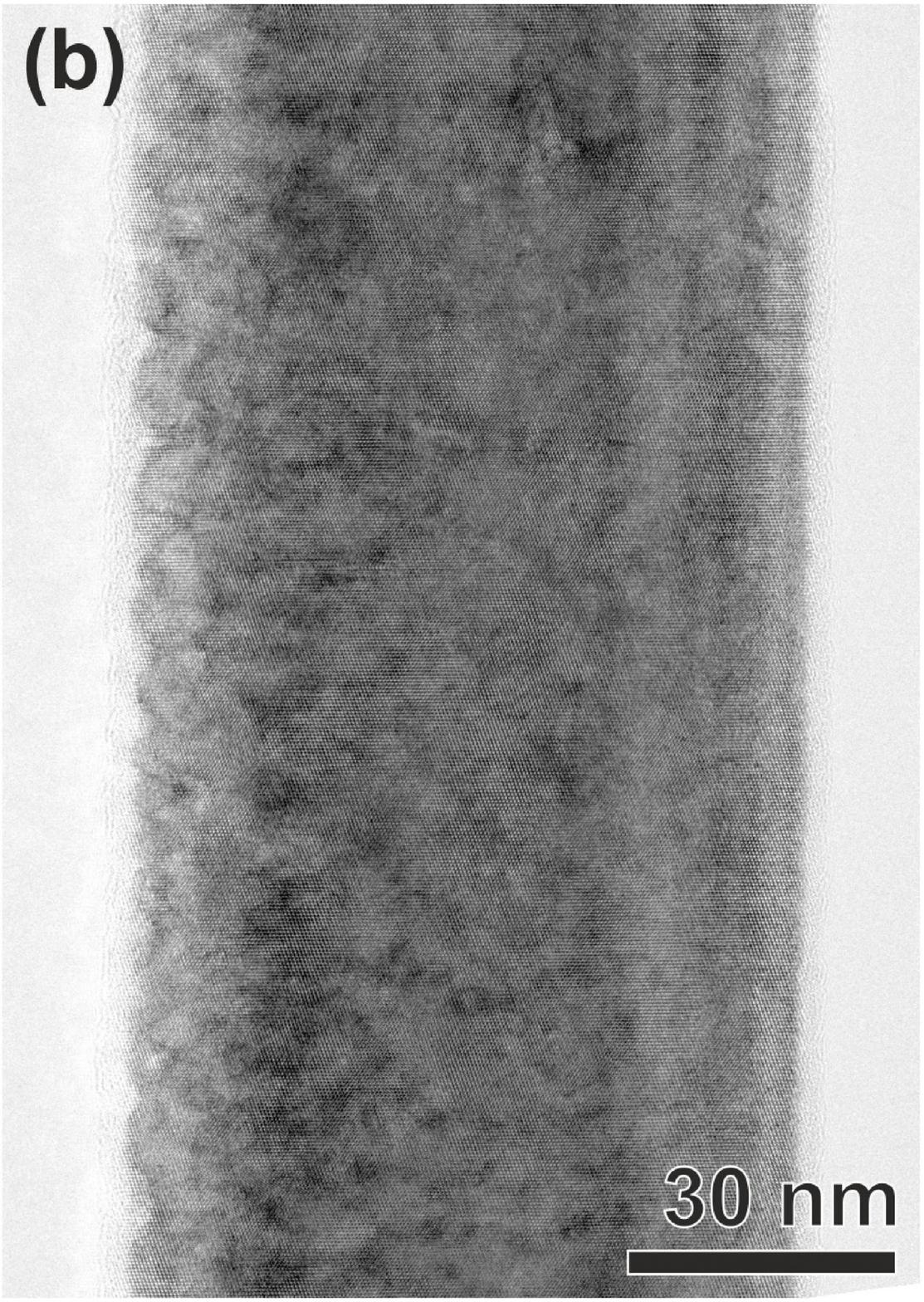}
		\label{HRTEM_386}
	} \\
	\subfloat[]{
		\includegraphics[width=0.6\textwidth]{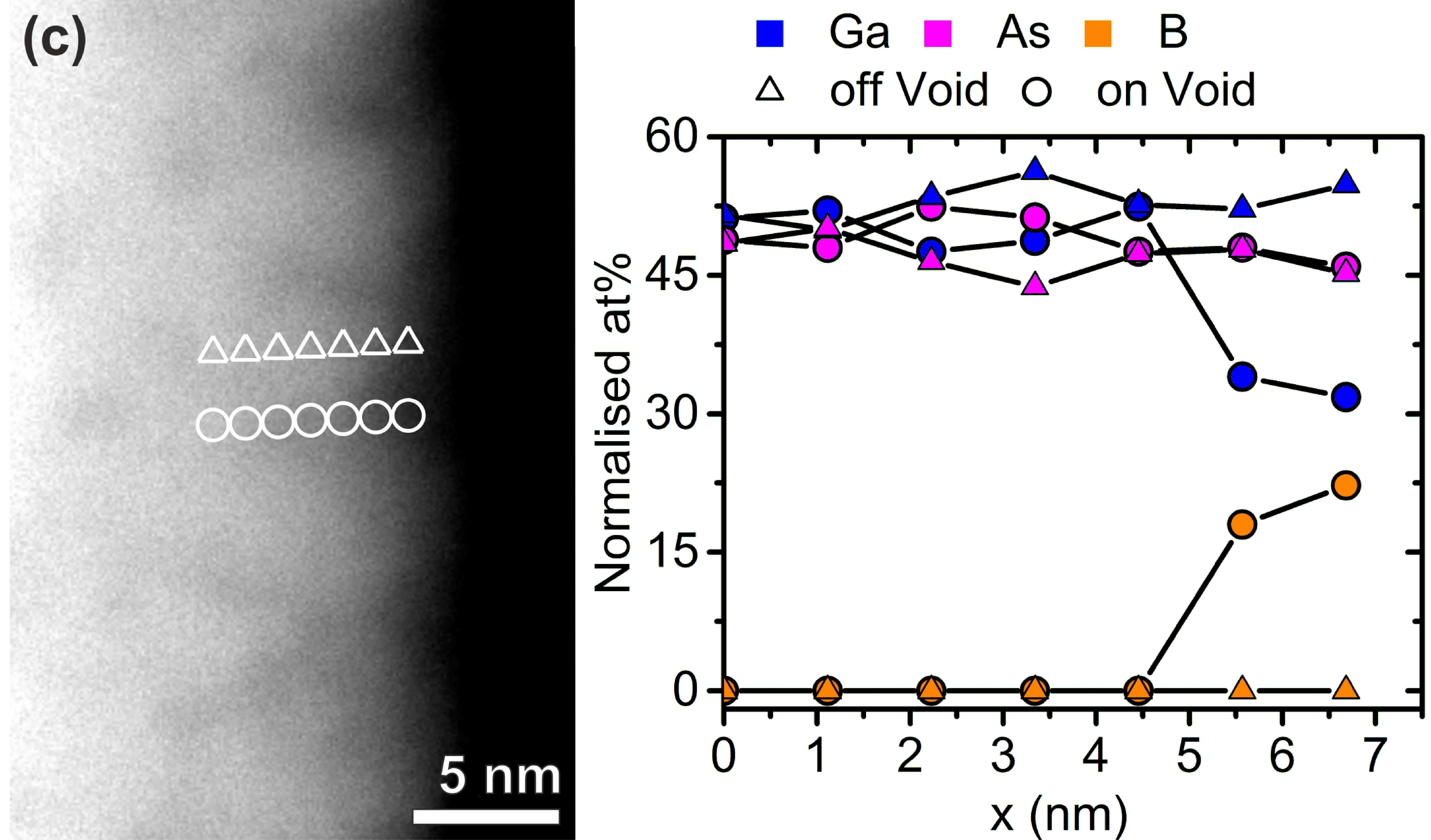}
		\label{EDXS}
	}
	\captionsetup{font=footnotesize}
	\caption{HRTEM images of nanowires from samples \protect\subref{HRTEM_380} C (with a zoom-in on one of the 'void' regions present at the NW surface, inset) and \protect\subref{HRTEM_386} E; (c) EDXS measurements on one of the spots visible in (a), and in an adjacent region. The measurement positions are marked in the HAADF images, left with the respective symbols used in the line scan, right.}
	\label{TEM}
\end{center} 
\end{figure*}

\section{Results \& Discussion}
Figure \ref{lengthstats} shows the average NW length versus the nominal boron concentration, as calculated from a set of 50-70 NWs per sample, for samples A1-E. It is immediately clear that the final nanowire length is inversely proportional to B flux, regardless of sample growth time. The growth rate is expected to be nonlinear at the start of the growth \cite{ramdani2012arsenic}, and the lengths of samples A1 (5.7$\mu$m after 60 mins) and A2 (1.3$\mu$m after 21 mins) confirm this. Assuming a linear growth rate starting from 15 mins of growth, we can estimate that the length of the GaAs stems after 18 minutes of growth should be $\sim$ 960nm. By comparing samples grown for a total of 60 minutes (B, D) or 48 minutes (C, E), we can conversely conclude that the radial growth (plotted in figure \ref{diamstats}) is dependent on both B content and growth time; both an increased B flux (i.e. sample B versus sample D, sample C versus sample E) and longer growth (i.e. samples B and D versus samples C and E) lead to enhanced radial growth, and to increased ratio of the base/tip diameter, indicating wire tapering. Since axial and radial growth depend on the local density of NWs, the number of wires per $\mu$m$^2$ was calculated for several substrate positions for each sample. This is plotted in figure \ref{densitystats}. There is generally good agreement in NW densities between samples, and density fluctuations which do occur are attributed to differences in surface morphology between wafers. Additionally, the variations in density can not explain the significant differences in NW morphology demonstrated in figures \ref{lengthstats} and \protect\subref*{diamstats}. 

Typical TEM images of the tips of nanowires grown with and without B are shown in figure \ref{droplets}. Figure \ref{droplets}a is taken from a pure GaAs NW sample grown under similar conditions. Without boron, the droplet remains intact and has a contact angle of 130\degree, which is typical for self-catalyzed GaAs nanowires and allows for continued axial growth and minimal tapering \cite{cirlin2010self}. In contrast, when B is introduced to the growth we can see that the droplet crystallises, with a final contact angle of around 30\degree, and the crystal structure of the wire closely follows that which we would expect from droplet consumption after NW growth \cite{krogstrup2011impact}. A first-principles B-Ga phase diagram indicates that the droplet under boron flux should still be composed entirely of gallium \cite{okamoto2015supplemental}, and indeed EDXS mapping of a NW tip indicates that the B level is below the detection limit. Unintentional consumption of the Ga droplet would prevent axial growth, and material may instead be deposited on the sidewalls, leading to radial NW growth. 

In order to understand the suppression of the axial nanowire growth, and simultaneous enhancement of the radial growth, we must look at the kinetics of self-catalyzed GaAs nanowire growth. Under normal conditions, Ga is supplied from a droplet, which is sustained by the impinging beam flux and sidewall diffusion of Ga \cite{dayeh2009surface, huang2010control}, while As is supplied via direct impingement and reemission from the substrate and neighbouring nanowires \cite{ramdani2012arsenic}. It is well-known that the Ga droplet can be consumed by stopping the Ga flux and supplying either high or low As$_4$ flux \cite{dastjerdi2016methods}, and that during consumption the droplet contact angle to the NW changes as the Ga supply dwindles. This change in contact angle leads to a change in the droplet-solid surface energy and the droplet supersaturation, and subsequently a change in crystal structure \cite{glas2007why} and therefore droplet consumption in a zinc-blende (ZB) NW leads to a `signature' crystal structure composition as the NW changes to wurtzite (WZ) structure and back, with the transitions between crystal structures marked by stacking faults (SFs) \cite{yu2012evidence}. 

The previously reported segregation of B in GaAs \cite{dumont2003surface} suggests that B acts as a surfactant. In our HRTEM and HAADF images a defect rich NW surface is visible, which leads to a speckled contrast in projection (figures \ref{HRTEM_380} and b). The defects at the nanowire edges correlate with small reductions in wire diameter, leading to 'voids' along the nanowire length. These voids present with an inverse pyramidal shape and clearly
lead to the dark spots visible around the whole NW, as can be seen in the inset to figure 4a.To determine whether there is also a composition change in the nanowire material, EDXS linescans were taken on one of these voids, and on an adjacent region (figure 4c). There is a significant B signal at the position of the void, and a concurrent reduction in Ga, despite a nominal B content \textless1\% (table \ref{table1}) which is below the detection limit (therefore, it is impossible to quantify via EDXS the amount of B incorporated on the lattice). This suggests segregation of BAs or some sub-arsenide, leading to alternating B- and Ga-rich regions on the nanowire surface. These voids on the nanowire surface are more numerous in samples grown at higher B flux (figure 4b), with an imageJ analysis indicating that near the nanowire tip, these boron-rich regions covered 8.9\% (14\%) on samples C (E) respectively. This indicates an increase in the surface coverage of B as the B flux is increased. The impact of increasing the growth time was not yet investigated structurally, but is also expected to increase surface coverage of B. It should be noted that an asymmetry in these voids was sometimes observed, as visible in figure 4b, which we attribute to a beam shadowing effect due to the relatively high wire density (see figure \ref{SEM2}), offering further evidence that these high-B regions arise due to sidewall nucleation rather than through VLS growth. 

A final point to consider is the increased tapering visible in SEM images (figure \ref{SEM}) and statistically confirmed to increase with both boron concentration and growth time (figure \ref{diamstats}). NW tapering was also observed in B-doped Ge nanowires \cite{Fukata2010doping} grown via chemical vapor deposition (CVD). Furthermore, a lower axial growth rate was found in highly B-doped Si NWs grown by CVD \cite{Pan2005Effect}. That these effects also occur due to the addition of boron during MBE growth of GaAs is interesting, as in this case the boron should be expected to incorporate substitionally as well as on antisite defects, and analysing the NW growth allows us to further elucidate some of the issues which occur in planar growth of B$_x$Ga$_{1-x}$As by highlighting which growth parameters are most affected by the presence of boron, as described below.

Massies and Grandjean \cite{massies1993surfactant} found that surfactants could either increase or decrease surface diffusion in GaAs; moreover, this was related to whether the atoms occupied interstitial surface sites (leading to an increased diffusion length) or substitutional sites (leading to a reduced diffusion length). In the case of substitutional surfactants, Ga adatoms adsorbed on the NW surface may be exchanged with the surfactant species (in this case B) and incorporated subsurface, where the energy barrier for migration involves breaking all their existing bonds.  A reduction in diffusion length could follow from the surface coverage of boron, and be further influenced by the increasing surface roughness clearly visible in figures \ref{TEM}\subref*{HRTEM_380} and b. A decreased adatom diffusion length could further come from the anomalous bond behaviour of B compared to other group III elements \cite{wentzcovitch1986theory,aldridge2011group, bouhafs2000trends}. 

 \subsection{Modelling of adatom diffusion length under B flux} 
From the net influx of Ga atoms in the droplet during VLS growth \cite{priante2013stopping} we can derive the rate of change of Ga atoms in the droplet as:
\begin{eqnarray}
\frac{dN_{Ga}}{dt} = J_{Ga,eff}+ J_{Ga}\left(\sin(\alpha_{Ga})2R\lambda_f + \cos(\alpha_{Ga})\pi(\lambda_s^2 - R^2)\right) \left(1-\frac{\theta_l}{\theta_f}\right) \nonumber \\
 - \frac{\pi R^{2}}{\Omega_{GaAs}} \frac{dL}{dt} \label{eq1} 
\end{eqnarray}

where J$_{Ga,eff}$ is the effective flux impinging on the Ga droplet, which takes into account the droplet shape following reference \cite{glas2010vapor}; J$_{Ga}$ is the direct Ga beam flux; $\alpha_{Ga}$ is the angle of the Ga beam on the droplet; R is the nanowire radius; $\lambda_f$ and $\lambda_s$ are the diffusion lengths of Ga adatoms on the nanowire sidewalls and the Si(111) substrate, respectively; and $\theta_l/\theta_f$ is the reverse diffusion flux or the flux of Ga adatoms to the droplet, which is given by the ratio of gallium activities in the liquid phase, $\theta_l$, and on the nanowire sidewalls, $\theta_f$. The $\lambda_s^2 - R^2$ term refers to adatoms reaching the droplet from the substrate, and is only taken into consideration if $\lambda_f > L$, where L is the nanowire length, i.e. adatoms reaching the NW base will also reach the droplet. The final term on the right-hand side corresponds to atoms from the droplet which are incorporated in the NW during axial growth, where $\frac{dL}{dt}$ is the NW growth rate, and $\Omega_{GaAs}$ is the elementary volume of Ga in solid GaAs (0.0451nm$^3$ in ZB \cite{dubrovskii2008growth}). If the number of Ga atoms in the droplet is decreasing, this elongation `sink' must be greater than the net influx of Ga atoms to the droplet, therefore:
\begin{eqnarray}
 J_{Ga}\left(\sin(\alpha_{Ga})2R\lambda_f + \cos(\alpha)\pi(\lambda_s^2 - R^2)\right)\left(\frac{\theta_l}{\theta_f}\right) + \frac{\pi R^{2}}{\Omega_{GaAs}}\frac{dL}{dt} > \nonumber \\
 J_{Ga,eff} +  J_{Ga}\left(\sin(\alpha_{Ga})2R\lambda_f + \cos(\alpha)\pi(\lambda_s^2 - R^2)\right)  \label{eq2}
\end{eqnarray}

After opening the boron shutter, the Ga flux stays constant, and from available B-Ga phase diagrams \cite{okamoto2015supplemental}, B is insoluble in Ga at our growth temperature, so we assume that the droplet is still composed of Ga. Therefore, we assume that $J_{Ga}$, $J_{Ga,eff}$ and $\alpha_{Ga}$ do not change. The radius R can also be considered constant in the moments after opening the boron shutter. 

The gallium activities in the droplet and on the sidewalls are given by \cite{dubrovskii2009gibbs}:
\begin{eqnarray}
\theta_l =\exp\left(\frac{\mu_l^\infty}{k_B T} + \frac{R_{GT}}{R}\right) \label{eq3} \\
\theta_f = J_{Ga} \tau_f \sigma_f \sin(\alpha_{Ga}) \label{eq4}
\end{eqnarray}

where $\mu_l^\infty$ is the infinite chemical potential of Ga, $k_B$ is Boltzmann's constant, T is the growth temperature, $R_{GT}$ is the Gibbs-Thomson radius, $\tau_f$ is the surface lifetime of Ga adatoms on the sidewalls (see below), and $\sigma_f$ is the element adsorption area of Ga on the nanowire sidewalls (0.45nm$^2$). Many of these terms are constant; as above, if the droplet composition does not change, $\mu_l^\infty$ should not change. R$_{GT}$ depends on the composition of the droplet, calculated as in \cite{dubrovskii2009gibbs}, and as above is considered to be pure Ga. Finally we are left with the surface lifetime of Ga adatoms, which is related to the sidewall diffusion length via:

\begin{equation}
\lambda_f = \sqrt{D_f \tau_f} \label{eq5}
\end{equation}

where D$_f$ is the diffusion constant of Ga on GaAs (1.38e-14$m^2s^{-1}$). Combining equations \ref{eq2} and \ref{eq5} we see that the left-hand term in \ref{eq1} has a $1/\sqrt{\tau_f}$ dependence while the second term on the right depends on $\sqrt{\tau_f}$. Therefore, a reduction in the Ga adatom lifetime (thus reduced diffusion length) would lead to a shrinking right-hand term and an expanding left-hand term, and eventually to a negative $dN_{Ga}/dt$. This would subsequently decrease the droplet volume and the contact angle $\beta$, which has been shown via the change in surface energies to lead to the crystal phase switch observed in our samples \cite{glas2007why}. 

We performed a model of the nanowire growth during droplet consumption, taking into account diffusion on the NW sidewalls as well as a contribution from substrate diffusion, since the NW height upon opening the B shutter is small. Here we calculate the change in nanowire length given by \cite{dubrovskii2009gibbs}:
\begin{equation}
\frac{dL}{dH} = \frac{\theta_s - \theta_l}{\theta_v} + \Bigg(\frac{dL}{dH}\Bigg)_{diff}, \label{eq6}
\end{equation}

with 
\begin{equation}
\Bigg(\frac{dL}{dH}\Bigg)_{diff} = \frac{2\lambda_s}{R}\Bigg[\frac{\delta(1-\frac{\theta_l}{\theta_s}) + (\frac{\lambda_f}{\lambda_s})\tan\alpha_{Ga}(1-\frac{\theta_l}{\theta_f})U(\frac{L}{\lambda_f})}{U'(\frac{L}{\lambda_f})}\Bigg] \label{eq7}
\end{equation}

Where dH/dt is the rate of deposition, given from the equivalent layer thickness growth rate (in this case, 0.1$\mu m/hr$), $\theta_s$ is given by $\theta_{s} = J\tau_{s}\sigma_{s}\sin\alpha$ and represents the adatom coverage of the substrate, with parameters analogous to those previously discussed for the sidewalls; $\theta_v$ is the activity of the vapor phase, calculated assuming $\frac{\theta_v}{\theta_s} = 2.11$ \cite{dubrovskii2009gibbs, dubrovskii2008growth}; $\lambda_s$ is the diffusion length of Ga adatoms on the Si(111) substrate; $\delta = K_1(\frac{R}{\lambda_s})/K_0(\frac{R}{\lambda_s})$ where $K_i$ are the modified Bessel functions of the second kind; and the functions $U(x)=\sinh(x)+\nu\delta[\cosh(x)-1]$ and $U'(x)=\cosh(x)+\nu\delta \sinh(x)$ with $\nu = D_s\lambda_f\sigma_f/D_f\lambda_s\sigma_s$. 

\begin{figure}[!t]
\begin{center}
	\subfloat[]{
		\includegraphics[height=5cm]{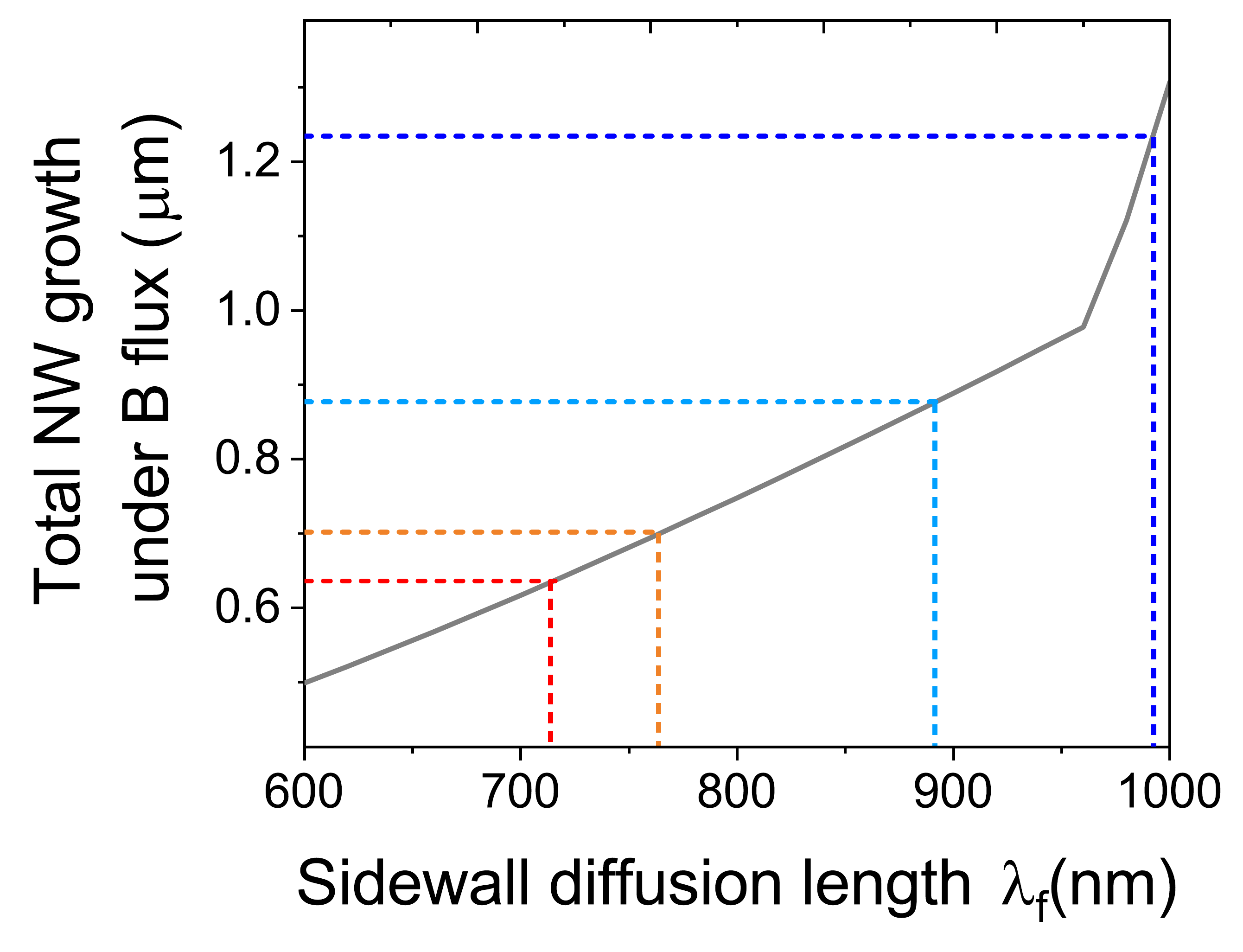}
		\label{diffmodel}
	}
	\hspace{0.5cm}
	\subfloat[]{
		\includegraphics[height=5cm]{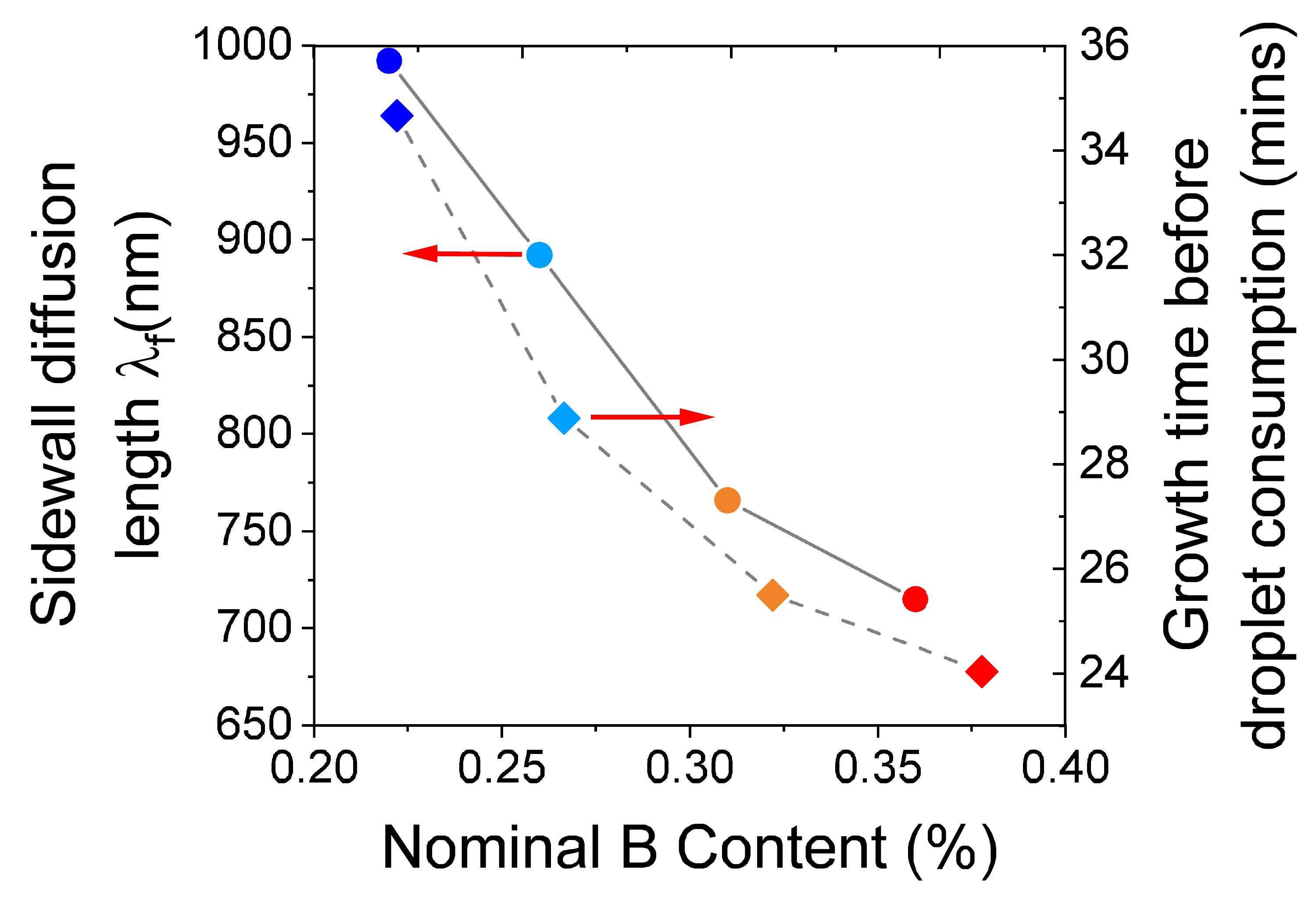}
		\label{diffmodel_lengths}
	}
	\captionsetup{font=footnotesize}
	\caption{\protect\subref{diffmodel} Simulated NW growth against sidewall diffusion length for a NW of 28nm radius. Dashed lines indicate the average NW growth calculated for samples B-E, and the corresponding diffusion length. The kink past $\lambda_f > 960nm$ comes from the contribution of substrate adatoms to the droplet. \protect\subref{diffmodel_lengths} Diffusion length calculated from \protect\subref{diffmodel} and growth time under B flux before droplet consumption, for samples B-E.}
	\label{diff}
\end{center} 
\end{figure}

For varying $\lambda_s$ and $\lambda_f$, equations \ref{eq1} and \ref{eq6} are solved iteratively and the change in NW length, droplet volume and droplet angle are modelled in the time after the boron shutter is opened, until the value of the contact angle reaches 30\degree, which is measured to be the angle of the droplet after consumption, or until the maximum growth time is reached if the droplet is not consumed. The value of $\lambda_s$ is constrained by the contribution of substrate adatoms to the droplet, since both small and large values of $\lambda_s$ lead to an expanding droplet.In the case of a high surface diffusion length, this is due to an increased 
flux to the droplet; in the case of a low $\lambda_s$, the axial growth rate is reduced, and the constant fluxes to the droplet are not fully balanced by the 'sink' of atoms being incorporated into the growing NWs. Considering these constraints, we found a value of $\lambda_s$ of 340nm, which is slightly lower but comparable to the substrate diffusion length found in a previous study \cite{detz2017lithography}. 

Figure \ref{diffmodel} shows the total nanowire growth under boron flux for different values of $\lambda_f$, for an average NW radius of 28nm. The pronounced kink at $\lambda_f$ \textgreater 960nm comes from the contribution of substrate adatoms to the droplet, past which the third time in equation \ref{eq1} becomes relevant. This model allows us to estimate the sidewall diffusion length for each sample, plotted in figure \ref{diffmodel_lengths}. In all cases, this corresponds to the droplet being consumed before the end of growth, which is consistent with HRTEM images, and explains the independence of NW length on the growth time. In addition, since the diffusion length is less than the final NW height in all cases, this explains the enhanced base/tip tapering ratio of NWs grown under B flux. These diffusion lengths can be compared to the typical sidewall diffusion length during MBE growth of pure GaAs nanowires, which is on the order of several microns \cite{dubrovskii2009gibbs}. 

\begin{figure}[!b]
\begin{center}
	\subfloat[]{
		\includegraphics[width=0.4\textwidth]{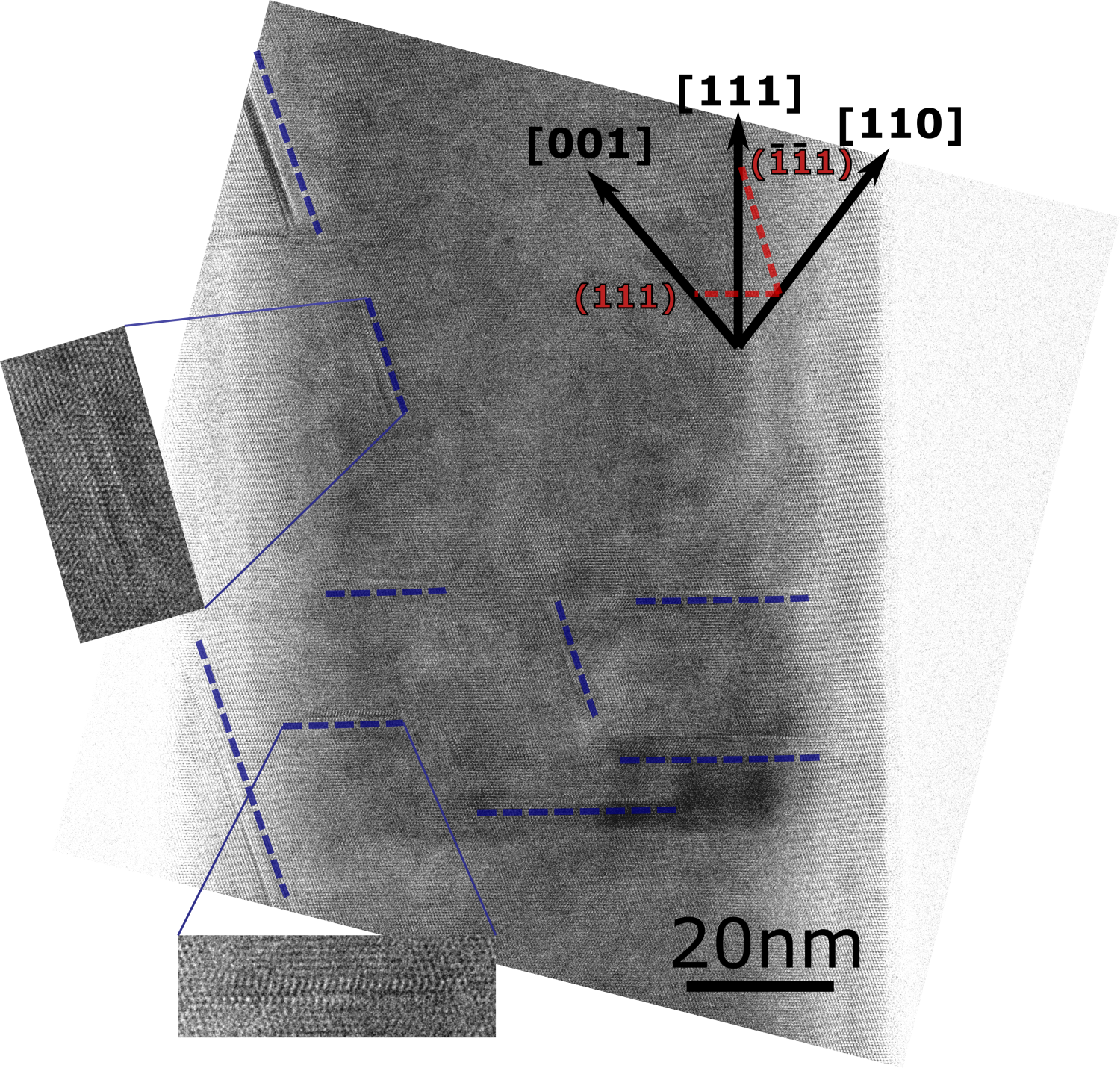}
		\label{AlGaAs_HRTEM}
	}
	\hspace{0.5cm}
	\subfloat[]{
		\includegraphics[width=0.37\textwidth]{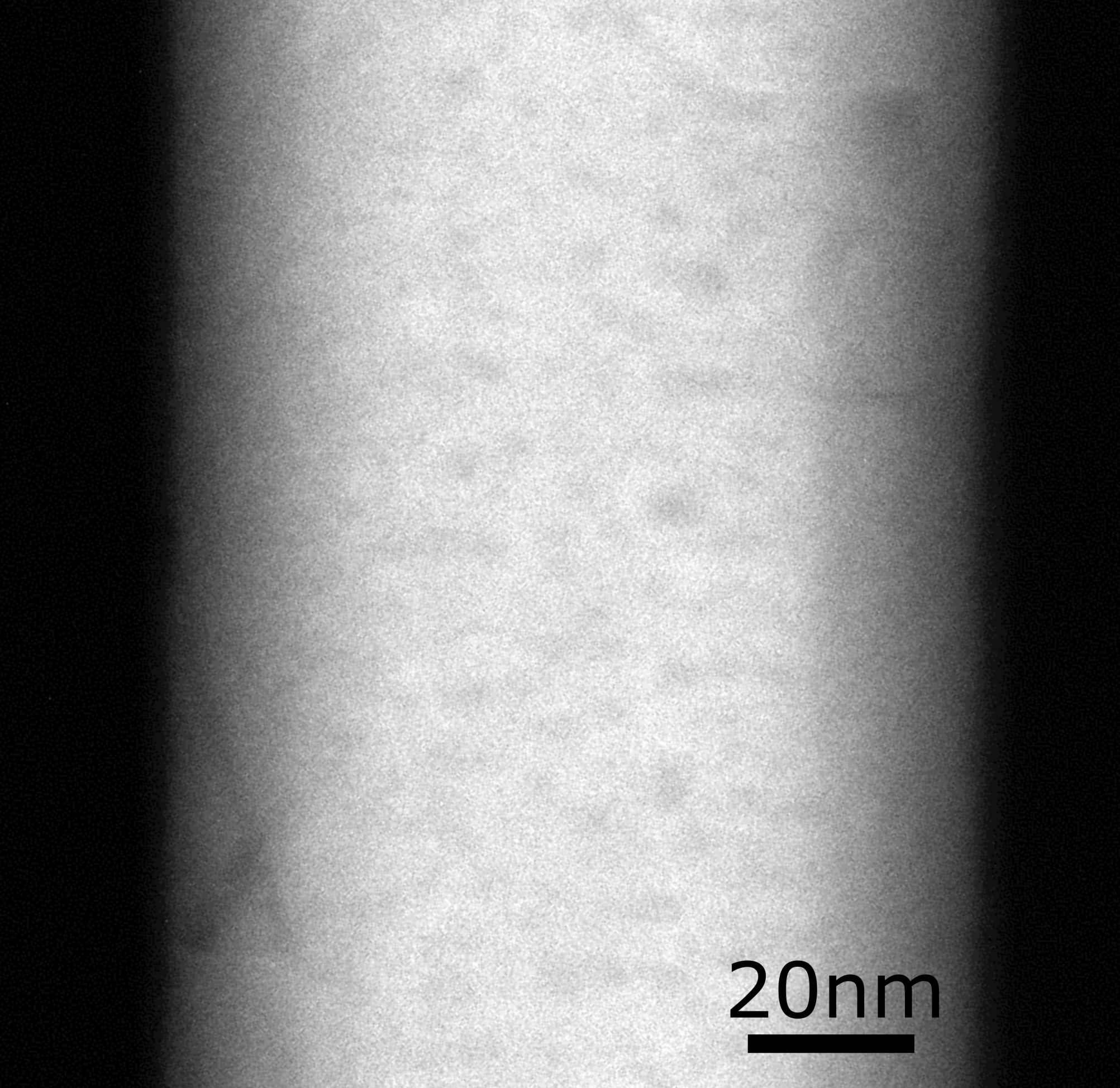}
		\label{AlGaAs_HAADF}
	}
	\captionsetup{font=footnotesize}
	\caption{\protect\subref{AlGaAs_HRTEM} HRTEM image of GaAs/(B)GaAs/AlGaAs core-multishell nanowire. Blue dashed lines indicate shell defects parallel to the (111) and ($\overline{1}\overline{1}$1) planes. \protect\subref{AlGaAs_HAADF} HAADF image of GaAs/(B)GaAs/AlGaAs core-multishell nanowires, where B-rich areas of contrast are clearly visible, alongside similar defects to those highlighted in \protect\subref{AlGaAs_HRTEM}.}
	\label{AlGaAs}
\end{center} 
\end{figure}

\subsection{Radial overgrowth of AlGaAs}
The (B)GaAs shell material is highly crystalline and follows the crystal structure of the NW core, as indicated by the ZB segment shown in figure 3b, although the segregation of B leads to a rough surface morphology. Nonetheless, the epitaxial quality of the material was further confirmed by the overgrowth of an Al$_{0.42}$Ga$_{0.58}$As shell (figure \ref{AlGaAs_HRTEM}), which was grown for 12 minutes and capped with GaAs to avoid oxidation. The AlGaAs grew conformally and with a smooth surface, despite the same B-rich areas of contrast being visible at the (B)GaAs-AlGaAs interface (figure \ref{AlGaAs_HAADF}). 

In figure \ref{AlGaAs_HRTEM} it is clear that the overgrowth of AlGaAs leads to the formation of characteristic defects which can be identified as stacking faults parallel to the (111) and ($\overline{1}\overline{1}$1) planes. The combination of growth on the rough surface and the lattice mismatch between (B)GaAs and AlGaAs leads to the formation of stacking faults in the AlGaAs shell. HRTEM images indicate that these stacking faults are limited to the AlGaAs shell. We attribute this to the compressively strained growth of AlGaAs on the voids discussed earlier. Due to the additional faceting, perfect dislocations can dissociate here and form SFs.
  
\section{Conclusion}
In conclusion, we have observed a strong impact of boron on the morphology of GaAs nanowires grown by solid-source MBE. Nanowires grown under B flux were found to be much shorter than bare GaAs NWs grown for the same length of time, and moreover the lengths reached by these wires depended solely on the B flux supplied during the second part of growth, irrespective of the total growth time. The NW shell material was found to have B-rich phases and a rough morphology, with coverage of B-rich areas on the surface increasing from 9\% to 14\% with increased boron content. From this we conclude that the surfactant nature of boron in GaAs leads to a reduction in Ga adatom surface lifetime, which limits the Ga supply to the droplet and eventually leads to droplet consumption and suppression of VLS growth. The surface lifetime reduction depends on the coverage of B and therefore on the B flux. 

These results offer insights into the growth of B$_{x}$Ga$_{1-x}$As, indicating that improving growth quality and boron content requires careful control over the growth parameters in order to reduce segregation of B and maximise adatom diffusion length. For example, it was found that the amount of substitutional B incorporated during planar growth drops sharply with growth temperatures $>$500\degree C \cite{ptak2012growth}, and that a high V/III ratio such as that used during NW growth increases B incorporation in the lattice \cite{groenert2004optimized}. This would suggest that, despite the difficulties of VLS growth incorporating B, the droplet could be intentionally consumed before growing a shell at a lower temperature for radial growth of BGaAs. This points to the possibility of incorporating B$_{x}$Ga$_{1-x}$As in radial heterostructures, especially as we were able to grow smooth, conformal AlGaAs shells on NWs despite their rough surface. Finally, the surfactant effect of adding boron during GaAs nanowire growth could be exploited, for example through engineering the growth on the NW sidewalls \cite{lewis2017self}, inducing crystal phase switching \cite{krogstrup2011impact}, or tuning the material properties through the surface roughness \cite{hochbaum2008enhanced}. \\

\textit{The authors would like to thank the Austrian Science Fund (FWF) projects P26100-N27 (H2N) and W1243 (Solids4Fun) for funding. This work was supported by the ESF under the project \\ CZ.02.2.69/0.0/0.0/16\_027/0008371.} \\

\textbf{This is the version of the article before peer review or editing, as submitted by the authors to Nanotechnology. IOP Publishing Ltd is not responsible for any errors or omissions in this version of the manuscript or any version derived from it. The Version of Record is available online at \url{https://doi.org/10.1088/1361-6528/aaf11e}}

\bibliographystyle{unsrt}
\bibliography{ref}

\begin{thebibliography}{10}

\bibitem{lauhon2002epitaxial}
Lincoln~J. Lauhon, Mark~S. Gudiksen, Deli Wang, and Charles~M. Lieber.
\newblock Epitaxial core–shell and core–multishell nanowire
  heterostructures.
\newblock {\em Nature}, 420(6911):57--61, 2002.

\bibitem{paladugu2007novel}
Mohanchand Paladugu, Jin Zou, Ya-Nan Guo, Graeme~J. Auchterlonie, Hannah~J.
  Joyce, Qiang Gao, H.~Hoe~Tan, Chennupati Jagadish, and Yong Kim.
\newblock Novel growth phenomena observed in axial {InAs}/{GaAs} nanowire
  heterostructures.
\newblock {\em Small}, 3(11):1873--1877, 2007.

\bibitem{caroff2009insb}
Philippe Caroff, Maria~E. Messing, B.~Mattias Borg, Kimberly~A. Dick, Knut
  Deppert, and Lars-Erik Wernersson.
\newblock {InSb} heterostructure nanowires: {MOVPE} growth under extreme
  lattice mismatch.
\newblock {\em Nanotechnology}, 20(49):495606, 2009.

\bibitem{hocevar2013residual}
Moïra Hocevar, Le~Thuy Thanh~Giang, Rudeesun Songmuang, Martien den Hertog,
  Lucien Besombes, Joël Bleuse, Yann-Michel Niquet, and Nikos~T. Pelekanos.
\newblock Residual strain and piezoelectric effects in passivated
  {GaAs}/{AlGaAs} core-shell nanowires.
\newblock {\em Applied Physics Letters}, 102(19):191103, 2013.

\bibitem{treu2013enhanced}
Julian Treu, Michael Bormann, Hannes Schmeiduch, Markus Döblinger, Stefanie
  Morkötter, Sonja Matich, Peter Wiecha, Kai Saller, Benedikt Mayer, and Max
  Bichler.
\newblock Enhanced {Luminescence} {Properties} of {InAs}–{InAsP}
  {Core}–{Shell} {Nanowires}.
\newblock {\em Nano letters}, 13(12):6070--6077, 2013.

\bibitem{liu2015coreshell}
Pengbo Liu, Hui Huang, Xueyu Liu, Min Bai, Danna Zhao, Zhenan Tang, Xianliang
  Huang, Ji-Yeun Kim, and Jinwei Guo.
\newblock Core–shell nanowire diode based on strain-engineered bandgap.
\newblock {\em physica status solidi (a)}, 212(3):617--622, 2015.

\bibitem{araki2013growth}
Yoshiaki Araki, Masahito Yamaguchi, and Fumitaro Ishikawa.
\newblock Growth of dilute nitride gaasn/gaas heterostructure nanowires on si
  substrates.
\newblock {\em Nanotechnology}, 24(6):065601, 2013.

\bibitem{Chen2014origin}
SL~Chen, Stanislav Filippov, Fumitaro Ishikawa, WM~Chen, and IA~Buyanova.
\newblock Origin of radiative recombination and manifestations of localization
  effects in gaas/ganas core/shell nanowires.
\newblock {\em Applied Physics Letters}, 105(25):253106, 2014.

\bibitem{gupta2000molecular}
V.~K. Gupta, M.~W. Koch, N.~J. Watkins, Y.~Gao, and G.~W. Wicks.
\newblock Molecular beam epitaxial growth of {BGaAs} ternary compounds.
\newblock {\em Journal of electronic materials}, 29(12):1387--1391, 2000.

\bibitem{wentzcovitch1986theory}
Renata~M. Wentzcovitch and Marvin~L. Cohen.
\newblock Theory of structural and electronic properties of {BAs}.
\newblock {\em Journal of Physics C: Solid State Physics}, 19(34):6791, 1986.

\bibitem{dumont2003surface}
H.~Dumont, D.~Rutzinger, C.~Vincent, J.~Dazord, Y.~Monteil, F.~Alexandre, and
  J.~L. Gentner.
\newblock Surface segregation of boron in {B} x {Ga} 1- x {As}/{GaAs} epilayers
  studied by {X}-ray photoelectron spectroscopy and atomic force microscopy.
\newblock {\em Applied physics letters}, 82(12):1830--1832, 2003.

\bibitem{detz2017growth}
H.~Detz, D.~MacFarland, T.~Zederbauer, S.~Lancaster, A.~M. Andrews, W.~Schrenk,
  and G.~Strasser.
\newblock Growth rate dependence of boron incorporation into {B} x {Ga} 1- x
  {As} layers.
\newblock {\em Journal of Crystal Growth}, 2017.

\bibitem{dheeraj2008zinc}
D.~L. Dheeraj, G.~Patriarche, L.~Largeau, H.~L. Zhou, A.~T.~J. van Helvoort,
  F.~Glas, J.~C. Harmand, Bjørn-Ove Fimland, and Helge Weman.
\newblock Zinc blende {GaAsSb} nanowires grown by molecular beam epitaxy.
\newblock {\em Nanotechnology}, 19(27):275605, 2008.

\bibitem{zhang2013self}
Yunyan Zhang, Martin Aagesen, Jeppe~V. Holm, Henrik~I. Jørgensen, Jiang Wu,
  and Huiyun Liu.
\newblock Self-catalyzed {GaAsP} nanowires grown on silicon substrates by
  solid-source molecular beam epitaxy.
\newblock {\em Nano letters}, 13(8):3897--3902, 2013.

\bibitem{ren2016new}
Dingding Ren, Dasa~L. Dheeraj, Chengjun Jin, Julie~S. Nilsen, Junghwan Huh,
  Johannes~F. Reinertsen, A.~Mazid Munshi, Anders Gustafsson, Antonius~TJ van
  Helvoort, and Helge Weman.
\newblock New insights into the origins of {Sb}-induced effects on
  self-catalyzed {GaAsSb} nanowire arrays.
\newblock {\em Nano letters}, 16(2):1201--1209, 2016.

\bibitem{jabeen2008self}
Fauzia Jabeen, Vincenzo Grillo, Silvia Rubini, and Faustino Martelli.
\newblock Self-catalyzed growth of {GaAs} nanowires on cleaved {Si} by
  molecular beam epitaxy.
\newblock {\em Nanotechnology}, 19(27):275711, 2008.

\bibitem{ramdani2012arsenic}
Mohammed~Reda Ramdani, Jean~Christophe Harmand, Frank Glas, Gilles Patriarche,
  and Laurent Travers.
\newblock Arsenic pathways in self-catalyzed growth of {GaAs} nanowires.
\newblock {\em Crystal Growth \& Design}, 13(1):91--96, 2012.

\bibitem{cirlin2010self}
GE~Cirlin, Vladimir~G Dubrovskii, Yu~B Samsonenko, AD~Bouravleuv, K~Durose,
  Yu~Yu Proskuryakov, Budhikar Mendes, L~Bowen, MA~Kaliteevski, RA~Abram,
  et~al.
\newblock Self-catalyzed, pure zincblende gaas nanowires grown on si (111) by
  molecular beam epitaxy.
\newblock {\em Physical Review B}, 82(3):035302, 2010.

\bibitem{krogstrup2011impact}
Peter Krogstrup, Stefano Curiotto, Erik Johnson, Martin Aagesen, Jesper
  Nyg{\textbackslash}a~ard, and Dominique Chatain.
\newblock Impact of the liquid phase shape on the structure of {III}-{V}
  nanowires.
\newblock {\em Physical review letters}, 106(12):125505, 2011.

\bibitem{okamoto2015supplemental}
H.~Okamoto.
\newblock Supplemental {Literature} {Review} of {Binary} {Phase} {Diagrams}:
  {Al}-{P}, {B}-{Ga}, {B}-{Nd}, {Ba}-{Ga}, {Bi}-{Cs}, {Ca}-{Ga}, {Cd}-{Gd},
  {Cr}-{Mo}, {Gd}-{Ni}, {Ni}-{Pb}, {Ni}-{Sc}, and {Sc}-{Sn}.
\newblock {\em Journal of Phase Equilibria and Diffusion}, 36(5):518--530,
  2015.

\bibitem{dayeh2009surface}
Shadi~A. Dayeh, Edward~T. Yu, and Deli Wang.
\newblock Surface diffusion and substrate- nanowire adatom exchange in {InAs}
  nanowire growth.
\newblock {\em Nano letters}, 9(5):1967--1972, 2009.

\bibitem{huang2010control}
Hui Huang, Xiaomin Ren, Xian Ye, Jingwei Guo, Qi~Wang, Xia Zhang, Shiwei Cai,
  and Yongqing Huang.
\newblock Control of the crystal structure of {InAs} nanowires by tuning
  contributions of adatom diffusion.
\newblock {\em Nanotechnology}, 21(47):475602, 2010.

\bibitem{dastjerdi2016methods}
M.~H.~T. Dastjerdi, J.~P. Boulanger, P.~Kuyanov, M.~Aagesen, and R.~R.
  LaPierre.
\newblock Methods of {Ga} droplet consumption for improved {GaAs} nanowire
  solar cell efficiency.
\newblock {\em Nanotechnology}, 27(47):475403, 2016.

\bibitem{glas2007why}
Frank Glas, Jean-Christophe Harmand, and Gilles Patriarche.
\newblock Why does wurtzite form in nanowires of {III}-{V} zinc blende
  semiconductors?
\newblock {\em Physical review letters}, 99(14):146101, 2007.

\bibitem{yu2012evidence}
Xuezhe Yu, Hailong Wang, Jun Lu, Jianhua Zhao, Jennifer Misuraca, Peng Xiong,
  and Stephan von Molnár.
\newblock Evidence for structural phase transitions induced by the triple phase
  line shift in self-catalyzed {GaAs} nanowires.
\newblock {\em Nano letters}, 12(10):5436--5442, 2012.

\bibitem{Fukata2010doping}
Naoki Fukata, Keisuke Sato, Masanori Mitome, Yoshio Bando, Takashi Sekiguchi,
  Melanie Kirkham, Jung-il Hong, Zhong~Lin Wang, and Robert~L Snyder.
\newblock Doping and raman characterization of boron and phosphorus atoms in
  germanium nanowires.
\newblock {\em ACS nano}, 4(7):3807--3816, 2010.

\bibitem{Pan2005Effect}
Ling Pan, Kok-Keong Lew, Joan~M Redwing, and Elizabeth~C Dickey.
\newblock Effect of diborane on the microstructure of boron-doped silicon
  nanowires.
\newblock {\em Journal of crystal growth}, 277(1-4):428--436, 2005.

\bibitem{massies1993surfactant}
J.~Massies and N.~Grandjean.
\newblock Surfactant effect on the surface diffusion length in epitaxial
  growth.
\newblock {\em Physical Review B}, 48(11):8502, 1993.

\bibitem{aldridge2011group}
Simon Aldridge and Anthony~J. Downs.
\newblock {\em The {Group} 13 {Metals} {Aluminium}, {Gallium}, {Indium} and
  {Thallium}: {Chemical} {Patterns} and {Peculiarities}}.
\newblock John Wiley \& Sons, 2011.

\bibitem{bouhafs2000trends}
Bachir Bouhafs, H.~Aourag, and M.~Certier.
\newblock Trends in band-gap pressure coefficients in boron compounds {BP},
  {BAs}, and {BSb}.
\newblock {\em Journal of Physics: Condensed Matter}, 12(26):5655, 2000.

\bibitem{priante2013stopping}
G.~Priante, S.~Ambrosini, Vladimir~G. Dubrovskii, A.~Franciosi, and Silvia
  Rubini.
\newblock Stopping and resuming at will the growth of {GaAs} nanowires.
\newblock {\em Crystal Growth \& Design}, 13(9):3976--3984, 2013.

\bibitem{glas2010vapor}
Frank Glas.
\newblock Vapor fluxes on the apical droplet during nanowire growth by
  molecular beam epitaxy.
\newblock {\em physica status solidi (b)}, 247(2):254--258, 2010.

\bibitem{dubrovskii2008growth}
VG~Dubrovskii, NV~Sibirev, JC~Harmand, and F~Glas.
\newblock Growth kinetics and crystal structure of semiconductor nanowires.
\newblock {\em Physical Review B}, 78(23):235301, 2008.

\bibitem{dubrovskii2009gibbs}
V.~G. Dubrovskii, N.~V. Sibirev, G.~E. Cirlin, I.~P. Soshnikov, W.~H. Chen,
  R.~Larde, E.~Cadel, P.~Pareige, T.~Xu, and B.~Grandidier.
\newblock Gibbs-{Thomson} and diffusion-induced contributions to the growth
  rate of {Si}, {InP}, and {GaAs} nanowires.
\newblock {\em Physical Review B}, 79(20):205316, 2009.

\bibitem{detz2017lithography}
Hermann Detz, Martin Kriz, Suzanne Lancaster, Donald MacFarland, Markus
  Schinnerl, Tobias Zederbauer, Aaron~Maxwell Andrews, Werner Schrenk, and
  Gottfried Strasser.
\newblock Lithography-free positioned gaas nanowire growth with focused ion
  beam implantation of ga.
\newblock {\em Journal of Vacuum Science \& Technology B, Nanotechnology and
  Microelectronics: Materials, Processing, Measurement, and Phenomena},
  35(1):011803, 2017.

\bibitem{ptak2012growth}
AJ~Ptak, DA~Beaton, and A~Mascarenhas.
\newblock Growth of bgaas by molecular-beam epitaxy and the effects of a
  bismuth surfactant.
\newblock {\em Journal of Crystal Growth}, 351(1):122--125, 2012.

\bibitem{groenert2004optimized}
ME~Groenert, R~Averbeck, W~H{\"o}sler, M~Schuster, and H~Riechert.
\newblock Optimized growth of bgaas by molecular beam epitaxy.
\newblock {\em Journal of crystal growth}, 264(1-3):123--127, 2004.

\bibitem{lewis2017self}
Ryan~B. Lewis, Pierre Corfdir, Jesús Herranz, Hanno Küpers, Uwe Jahn, Oliver
  Brandt, and Lutz Geelhaar.
\newblock Self-assembly of {InAs} nanostructures on the sidewalls of {GaAs}
  nanowires directed by a {Bi} surfactant.
\newblock {\em arXiv preprint arXiv:1704.08014}, 2017.

\bibitem{hochbaum2008enhanced}
Allon~I. Hochbaum, Renkun Chen, Raul~Diaz Delgado, Wenjie Liang, Erik~C.
  Garnett, Mark Najarian, Arun Majumdar, and Peidong Yang.
\newblock Enhanced thermoelectric performance of rough silicon nanowires.
\newblock {\em Nature}, 451(7175):163--167, 2008.

\end{thebibliography}





\end{document}